# Wide Field High Cadence CMB Survey Designs for Chilean Telescopes

Haruki Ebina,[1, 2, *] Reijo Keskitalo,[3, 2, 4] Julian Borrill,[3, 4] Steve K. Choi,[1, 5]
Theodore Kisner,[3, 2, 4] Sigurd Naess,[6] Michael Niemack,[1, 5] and Jason R. Stevens[7, 8]

[1]*Department of Physics, Cornell University, Ithaca, NY 14853, USA*
[2]*Department of Physics, University of California, Berkeley, CA 94720, USA*
[3]*Lawrence Berkeley National Laboratory, Berkeley, CA 94720, USA*
[4]*Space Sciences Laboratory, University of California, Berkeley, CA 94720, USA*
[5]*Department of Astronomy, Cornell University, Ithaca, NY 14853, USA*
[6]*Institute of Theoretical Astrophysics, University of Oslo,*
*P.O. Box 1029, Blindern, N-0315 Oslo, Norway*
[7]*University of Colorado, Boulder, Boulder, CO 80309, USA*
[8]*National Institute of Standards and Technology, Boulder, CO 80305, USA*

We present new wide field survey strategies for Chilean Large Aperture Telescopes (LAT) measuring the Cosmic Microwave Background (CMB), which we call Sinusoidal Modulated High Cadence Survey Strategies. The strategies were developed during the process of optimizing LAT measurements for the CMB-S4, Simons Observatory, and CCAT-prime collaborations. Observing more than $f_{sky} \sim 0.5$, the telescope consistently achieves high observation efficiency, even with Sun-avoidance enabled. Classical azimuthal scan survey strategies observing fields of equal size suffer from problems of observation depth non-uniformity relative to declination and lack of crosslinking. The new survey strategies described here significantly improve both uniformity and crosslinking while also enabling higher cadence observations for time-domain astrophysics. Uniformity and crosslinking are improved by modulation of azimuthal angular velocity and sinusoidal elevation nods, respectively. In particular, there is nearly uniform observation depth and crosslinking is improved from total lack of crosslinking near $-40°$ declination to clearing the strictest thresholds for crosslinking across the entire field. The simulated strategies are compared to the strategies used for the Atacama Cosmology Telescope and previously studied Simons Observatory survey strategies.

## I. INTRODUCTION

In observational cosmology, Cosmic Microwave Background (CMB) measurement remains one of the largest fields with next-generation experiments such as CMB Stage IV (CMB-S4, https://cmb-s4.org/), Simons Observatory (SO, https://simonsobservatory.org/), and CCAT-prime (http://www.ccatobservatory.org). These experiments in particular are equipped with a Chilean ground-based Large Aperture Telescope (LAT) [1][2]. Progressing towards these experiments, it is important to improve from previous experiments containing Chilean LATs in the construction of survey strategies.

Traditional scan strategies that have been used previously, such as for the Atacama Cosmology Telescope (ACT) consist of a series of constant speed azimuthal scans across the rising and setting sky, with observed fields summing up to $\sim 1400$sq. deg in the Advanced ACTPol (AdvACT) experiment [3]. While these strategies have been proved effective, they can be improved in many aspects, including the observation field area, observation depth uniformity and crosslinking, a parameter used to quantify the variety of scan angles through a point of sky. There has already been efforts made in some directions, such as those in Stevens et al. (2018) [4], which are used for the SO LAT forecasts [5]. While these strategies achieve wide-field coverage up to $\sim 33000$sq. deg., it does not achieve high cadence, as it aims to optimize observation depth relative to foreground contamination. As high cadence observation allows data collection for transient astrophysics, we aim to construct wide-field high cadence survey strategies. Although such high cadence survey strategies are already being considered within the community, informing the construct of CMB-S4 and CCAT-prime forecasts [6][7], this work will be their first presentation.

We present and analyze new survey strategies to address these points. We start with the Modulated High Cadence Survey Strategies (MHCSS) which addresses the observation field size, depth uniformity, and cadence. This is accomplished by constructing the strategies in a way that consistently observes the sky over a wide range of azimuth, with varying azimuthal scan rates to accommodate for geometric effects that leads to observation depth non-uniformity. Then we consider the Sinusoidal Modulated High Cadence Survey Strategies (SMHCSS), which is an improvement to the MHCSS with sinusoidal elevation nods added to create variety in scan angle. It is notable that these survey strategies are designed to achieve high cadence for time-domain astrophysics as well.

---

[*] he77@cornell.edu

For analysis, we employ a combination of TOAST (Time Ordered Astrophysics Scalable Tools) [8] and a set of individually developed software to create hitmaps, crosslinking maps, and boresight trace plots. We will use classical survey strategies used for the ACT measurements in data release 4 [9] and AdvACT [3], and previously studied Simons Observatory strategies [4] for comparison. In addition, we consider instrumental and strategic constraints to the telescope movement and verify that the SMHCSS is effective within these limits. Through our analysis we will observe that the SMHCSS provides high cadence wide-field survey strategies with uniform observation depth and satisfactory crosslinking for next-generation CMB experiments.

## II. MODULATED HIGH CADENCE SURVEY STRATEGIES

### A. Overview

The MHCSS are survey strategies designed to observe wide observation fields with uniform observation depth. This is done by employing wide azimuthal scans consistently, modulating the scan rate as a function of elevation and azimuth. It is worth noting that by design, these survey strategies will observe the same points in sky every 1-2 days. The form for velocity modulation is derived by setting the observation time $T(\theta)$ spent at a particular declination $\theta$ to be constant. With $\theta$ as declination, $\alpha$ as observation azimuth, $\beta$ as observation elevation, and $\theta_0$ as telescope latitude we perform the calculation as follows.

The declination $\theta$ of a point in sky is given by horizontal coordinates as

$$\sin\theta = \sin\theta_0 \sin\beta + \cos\theta_0 \cos\beta \cos\alpha \quad (1)$$

Using this, we can calculate $\frac{d\theta}{dt}$ as

$$\frac{d\theta}{dt} = \frac{\cos\theta_0 \cos\beta \sin\alpha}{\cos\theta} \frac{d\alpha}{dt} \quad (2)$$

Importantly, $\frac{d\theta}{dt} \propto T(\theta)^{-1}$, so setting $\frac{d\theta}{dt}$ to a constant $K$ gives us the desired modulation of telescope in horizontal coordinates. This is

$$\frac{d\alpha}{dt} = \frac{K \cos\theta}{\cos\theta_0 \cos\beta \sin\alpha} \quad (3)$$

Defining $\omega = \frac{d\alpha}{dt}$ as the azimuthal scan rate and setting the constant $\omega_0 = K/\cos\theta_0$ as the base scan rate, we

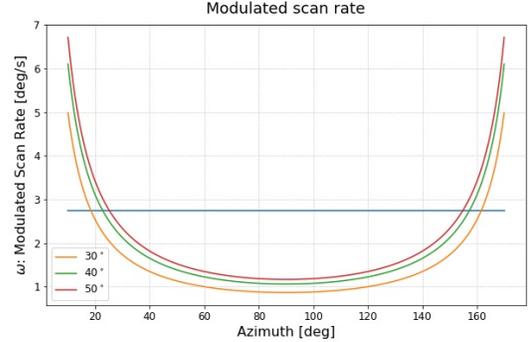

FIG. 1. Plot of the modulated azimuthal scan rate in °/s with respect to the azimuth coordinate of the telescope at 30, 40, 50° elevations. A base scan rate of 0.75°/s is used. The blue line indicates the desired maximum scan rate of 2.75°/s.

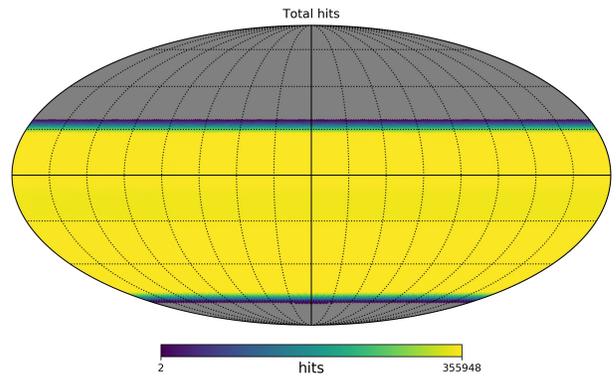

FIG. 2. A hitmap of a yearlong Modulated High Cadence Survey Strategy schedule observing at 40° elevation, [21°,159°] azimuth range and no Sun-Moon avoidance. Note the uniformity in observation depth.

obtain

$$\omega = \frac{\omega_0}{\cos\beta \sin\alpha} \quad (4)$$

An example of a modulated scan rate, with a base scan rate of 0.75°/s, is shown in Figure 1.

The effectiveness of these strategies is seen in an hitmap simulation using TOAST, shown in Figure 2. The efficiency of the survey strategy is 99.9%, as expected from the wide field and lack of Sun-Moon avoidance. Relative to the Advanced ACTPol (AdvACT) survey strategy observing $\sim$ 14000sq. deg. [3], as shown in Figure 3, the MHCSS observes $\sim$ 27000sq. deg. at 40° elevation and [21°,159°] azimuth range.

With a reasonable Sun-Moon avoidance angle of 45°, we will get the hitmaps shown in Figure 4. The left hitmap illustrates the map without modulation, while the right hitmap shows the map with modulation. Clearly, the modulation benefits in avoiding the concentration of observation time near the field edge, lowering the standard deviation of hits inside the field from 64400 to 23800 and raising the median hits inside the field from 228000 to 249000. It is also notable that the concentration of observation depth near the edge of the field in the hitmap without modulation is wasteful compared to having a similar concentration of observation depth near the center of the field. The histogram of hits for these hitmaps are shown in Figure 5, showing how the modulated survey strategy yields a higher median and lower standard deviation by the improved distribution of time among points on the sky. The observation efficiency for both survey strategies is 71.4%.

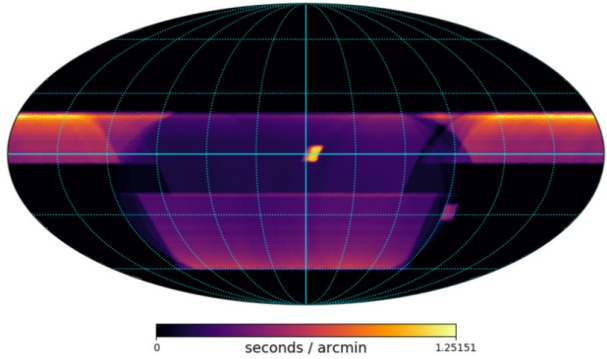

FIG. 3. A hitmap of a yearlong observation schedule for the AdvACT telescope. Note that the units shown here are not hits, but are seconds/arcmin, which is proportional to the number of hits.

### B. Constraints on scan rate and azimuthal range

MHCSS have constraints to their parameters from the maximum azimuthal velocity of the telescope and movement of sky after one azimuthal throw.

#### 1. Maximum Velocity

We set the high bound of telescope velocity to be at $2.75°/s$. In order to achieve this at a base scan rate of $0.75°/s$, we need to limit ourselves to the ranges shown in Figure 1.

Assuming an azimuthal range centered at $90°$, we require minimum azimuth and base scan rate shown in Figure 6 to stay under $2.75°/s$ at all points of the schedule.

#### 2. Movement in Sky

Considering the field of view of each optics tube, which is $1.3°$ in diameter for CCAT-prime and SO designs, we desire the movement in sky to be below $1.0°$ after one azimuthal throw.

Since the telescope will be pointing towards the same horizontal coordinates after one azimuthal throw, we can easily calculate the change in declination to be 0 through Eq. (1). Thus, it remains to calculate the change in right ascension. The right ascension of a point in sky can be described as

$$a = L - H \qquad (5)$$

where $a$ is the right ascension, $L$ is the Local Sidereal Time, and $H$ is the Local Hour Angle.

The local hour angle $H$ can be expressed through the equations

$$\cos H = (\sin \beta - \sin \theta \sin \theta_0)(\cos \theta \cos \theta_0)^{-1} \qquad (6)$$

$$\sin H = -\sin \alpha \cos \beta / \cos \theta \qquad (7)$$

where $\theta$ is declination, $\theta_0$ is observer latitude, $\beta$ is elevation, and $\alpha$ is azimuth. Similar to the case of declination, the change in $H$ after one azimuthal throw is 0 due to identical horizontal coordinates.

The calculation then reduces to the calculation of scan period, which we can calculate with the equation below

$$\Delta t = \frac{2}{x}\left(\left|\frac{d\alpha}{dt}\right|_{\alpha_{max}} + \left|\frac{d\alpha}{dt}\right|_{\alpha_{min}}\right) + 2\int_{\alpha_{min}}^{\alpha_{max}} \frac{\cos\beta \sin\alpha}{\omega_0}d\alpha \qquad (8)$$

where $\Delta t$ is change in (earth) time, $\alpha$ is azimuth, $\beta$ is elevation, $x$ is azimuthal turn-around acceleration, and $\omega_0$ is base scan rate. By simply correcting our units, we are done.

In the original settings - $[19°,161°]$ azimuth for $30°$ elevation, $[21°,159°]$ azimuth for $40°$ elevation, and $[26°,154°]$ azimuth for $50°$ elevation, all with base scan rate of $0.75°/s$ and azimuthal acceleration of $1.0°/s^2$ - we obtain a change in right ascension of $1.08°$, $0.95°$, and $0.78°$ respectively.

We can find the appropriate azimuthal ranges to achieve a $\Delta a$ of $0.5°$ and $1.0°$ with the same base scan rate and azimuthal acceleration. The results found are shown in Figure 7 and Table I.

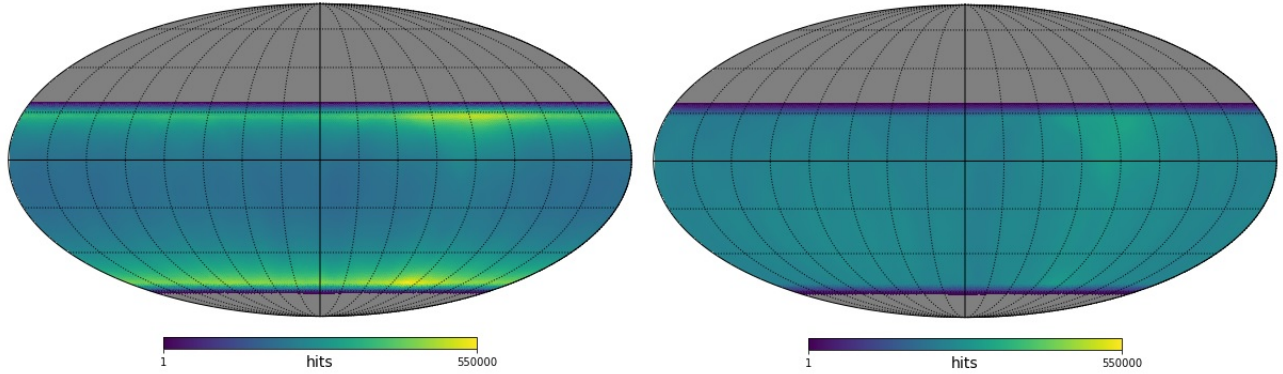

FIG. 4. A hitmap of a yearlong wide field observation schedule observing at 40° elevation, [21°,159°] azimuth range and 45° Sun-Moon avoidance angle. The left hitmap shows a schedule without modulated scan rate, while the right shows a schedule with modulated scan rate. Modulating the scan rate clearly improves the uniformity substantially, changing the standard deviation of hits inside the field from 64400 to 23800 and the median hits inside the field from 228000 to 249000.

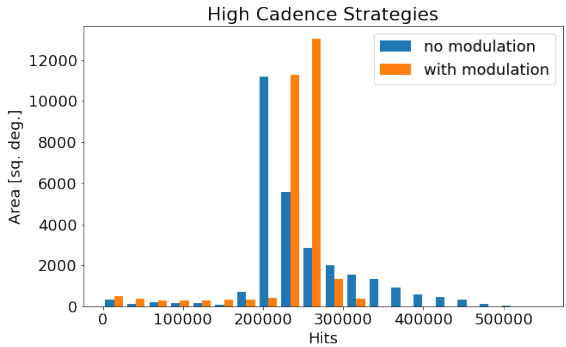

FIG. 5. A histogram of hits for the High Cadence Survey Strategy maps in Figure 4, with the map without velocity modulation in blue and the Modulated High Cadence strategy map in orange.

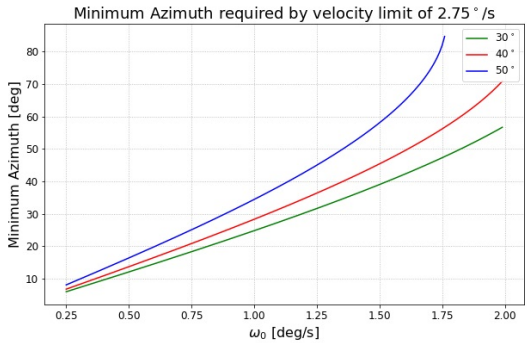

FIG. 6. The minimum azimuth, assuming an azimuthal range centered at 90°, that can be set by the maximum scan rate of 2.75°/s and the base scan rate shown on the x-axis. The angles in legend indicate the elevation of each scan.

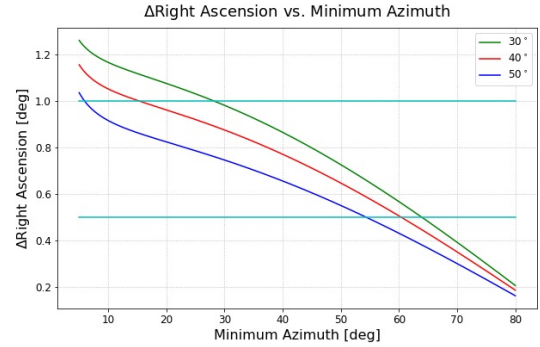

FIG. 7. The change in right ascension after one throw, plotted with respect to minimum azimuth in Modulated High Cadence Survey Strategies, assuming a symmetric azimuthal range about 90°. The horizontal lines illustrate the 0.5° and 1.0° criteria mentioned in the text. For comparison, the field of view diameter of a Simons Observatory and CCAT-prime optics tube is approximately 1.3°, although, with fewer detectors near the perimeter, more uniform map depth could be achieved by targeting the criteria mentioned above.

| $\Delta a$ [deg] | Elevation [deg] | Azimuthal range [deg] |
|---|---|---|
| 0.5 | 30 | [63.89,116.11] |
|  | 40 | [60.34,119.66] |
|  | 50 | [54.31,125.69] |
| 1.0 | 30 | [28.16,151.84] |
|  | 40 | [15.29,164.71] |
|  | 50 | [5.92,174.08] |

TABLE I. The azimuthal ranges at each elevation to achieve a 0.5° and 1.0° change in right ascension after one azimuthal throw in Modulated High Cadence Survey Strategies with a base scan rate of 0.75°.

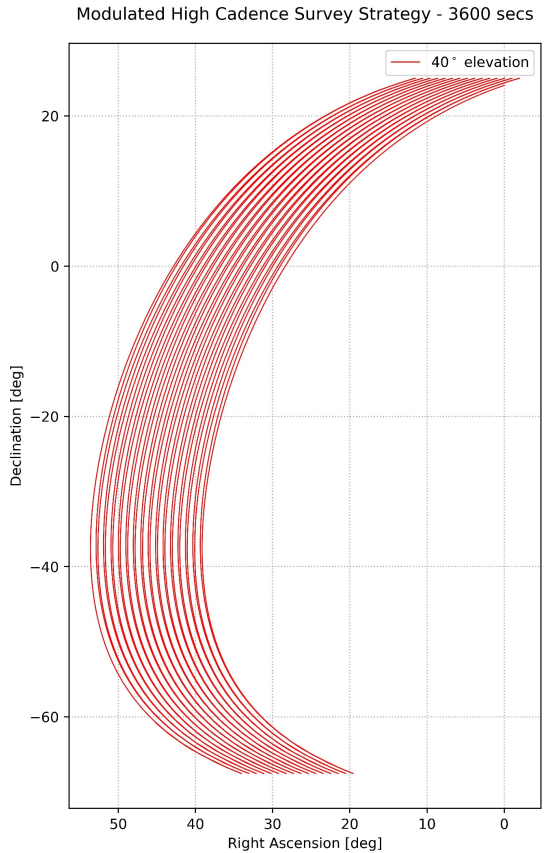

FIG. 8. The boresight traceo of a Modulated High Cadence Survey Strategy observing the rising sky at 0.75°/s base scan rate, 40° elevation and azimuthal range of [21°,161°].

### C. Boresight trace

In this section we will show boresight trace plots for MHCSS in equatorial coordinates to start our consideration of sinusoidal pattern scans.

We will construct the boresight trace plots by utilizing the equations in the previous section. The boresight trace plot for a MHCSS at 40° elevation and azimuthal range of [21°,161°] is shown in Figure 8. By zooming in, Figure 9 confirms that the plot constructed matches the movement of sky calculation of 0.95° above.

By plotting the scan pattern on the rising and setting sky together, as in Figure 10, we observe the clear lack of crosslinking, or variety in scan angle, near −40°

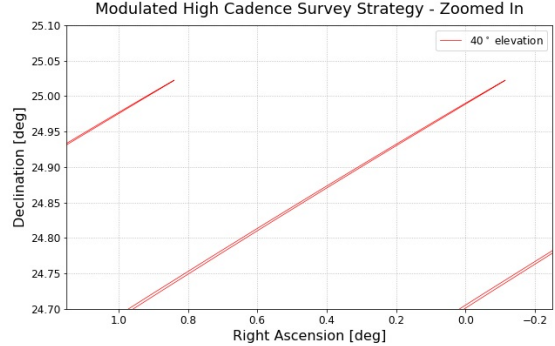

FIG. 9. A zoomed in picture of Figure 8, to illustrate the change in right ascension after one azimuthal throw. The change in the plot matches the calculation of 0.95°.

declination.

## III. SINUSOIDAL MODULATED HIGH CADENCE SURVEY STRATEGIES

Now we will move on to study the SMHCSS, which adds a sinusoidal movement in elevation to the MHCSS.

### A. Boresight Trace

We start by making boresight trace plots of the SMHCSS to illustrate the difference relative to the MHCSS. A simple boresight trace on the rising sky is shown in Figure 11. The effects of the sinusoidal modulation is easily visible when compared to Figure 8.

Similarly, we can contrast Figure 12 to Figure 10, where we plot the rising and setting sky scans of a SMHCSS and MHCSS together, respectively. It is clear that the crosslinking is enhanced through the sinusoidal modulations.

The boresight trace, in horizontal coordinates, is useful for gauging whether this strategy is realistic in the scope of telescope limitations. Figure 13 shows the position, velocity, acceleration, and jerk data for the SMHCSS with at 0.75°/s base scan rate, 40° elevation, 1° sinusoidal amplitude, 10s sinusoidal period, and azimuthal range of [21°,161°].

### B. Crosslinking Studies

To quantitatively analyze the advantages of the SMHCSS, we simulate the crosslinking map by largely modifying an existing algorithm, which quantifies crosslinking through a Stokes parameter approach [9][10].

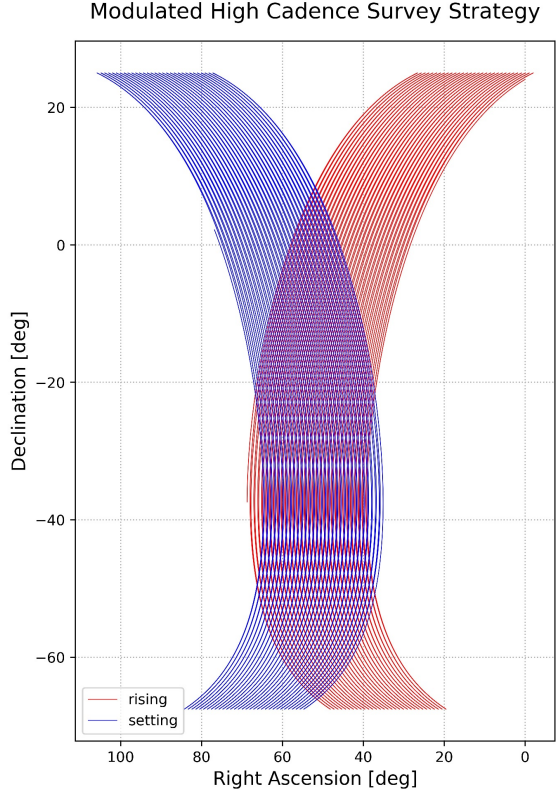

FIG. 10. The boresight trace plot for a Modulated High Cadence Survey Strategy observing the rising (red) and setting (blue) sky, at 0.75°/s base scan rate, 40° elevation and azimuthal range of [21°,161°]. Lack of crosslinking near −40° is clearly visible.

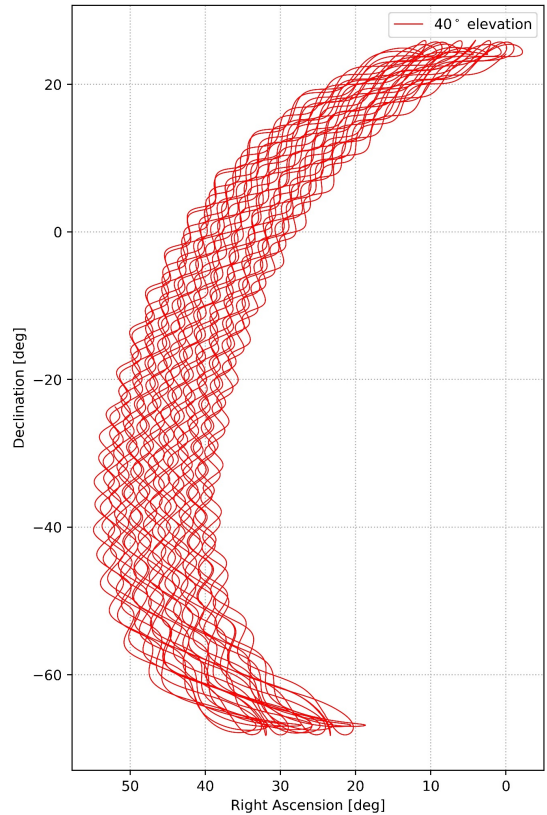

FIG. 11. The boresight trace plot for a Sinusoidal Modulated High Cadence Survey Strategy observing the rising sky at 0.75°/s base scan rate, 40° elevation, 1° sinusoidal amplitude and azimuthal range of [21°,161°].

Figure 14 shows the crosslinking maps using MHCSS and SMHCSS. Note that a crosslinking parameter of 0 corresponds to no crosslinking, while 1 corresponds to complete crosslinking. From the maps we find that while the MHCSS has nearly no crosslinking across a significant range of declinations, the points with the worst crosslinking in the SMHCSS map have a crosslinking parameter of 0.406, which is above the 0.3 value that Choi et al. (2020) [9] found to be preferential for CMB power spectrum analysis.

The sinusoidal oscillation period in Figure 14 is 11s. This is reasonable, as the time length of one azimuthal throw at 40° elevation and [21°,159°] azimuth range is ∼ 250s, which gives ∼ 20 periods in one throw (close to the number in Figure 11).

Figure 15 shows the effects of altering the period and amplitude of the sinusoidal pattern extensively.

## IV. CONCLUSION

In this work we present and explore new survey strategies, called the Sinusoidal Modulated High Cadence Survey Strategies (SMHCSS), for Chilean ground-based Large Aperture Telescopes (LAT) in next-generation Cosmic Microwave Background (CMB) observation experiments, such as CMB Stage IV (CMB-S4), Simons Observatory (SO), and CCAT-prime. SMHCSS are designed to improve from previous LAT survey strategies in terms of size of observation fields, uniformity in observation depth, and crosslinking. By design, these strategies also achieve high cadence, providing data applicable for time-domain astrophysics. Through the use of analytical tools such as observation hitmaps, boresight traces, and crosslinking maps, we show the advantage SMHCSS has to traditional LAT survey strategies in all three of the realms mentioned above. In particular, crosslinking saw drastic improvement as the entire map cleared the strictest threshold employed by Atacama Cosmology Telescope (ACT) data release 4 data analysis [9]. By showing that SMHCSS are able to clear constraints on maximum azimuthal velocity and movement of the sky after one scan throw we also verify that SMHCSS are realistic survey strategies to be implemented on next-generation experiments in the near future. Future steps to this study may investigate optimal combinations of sinusoidal oscillation amplitude and period to maximize crosslinking, as this work only contains preliminary effort in this direction, simply observing crosslinking at various sinusoidal amplitudes and periods.

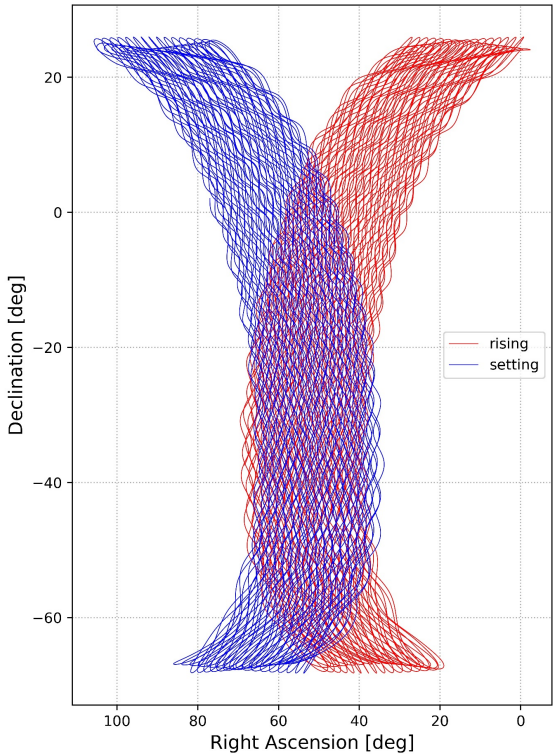

FIG. 12. The boresight trace plot for a Sinusoidal Modulated High Cadence Survey Strategy observing the rising (red) and setting (blue) sky at $0.75°$/s base scan rate, $40°$ elevation, $1°$ sinusoidal amplitude and azimuthal range of $[21°,161°]$.

## V. ACKNOWLEDGEMENTS

Support for this research was provided by the Department of Energy. MDN acknowledges support from NSF grant NSF grants AST-1454881 and AST-2117631. SKC acknowledges support from NSF award AST2001866. This research used resources of the National Energy Research Scientific Computing Center (NERSC), a U.S. Department of Energy Office of Science User Facility located at Lawrence Berkeley National Laboratory, operated under Contract No. DE-AC02-05CH11231 using NERSC award HEP-0021264. Some of the results in this paper have been derived using the `healpy` and `HEALPix`[1] packages.

---

[1] http://healpix.sf.net

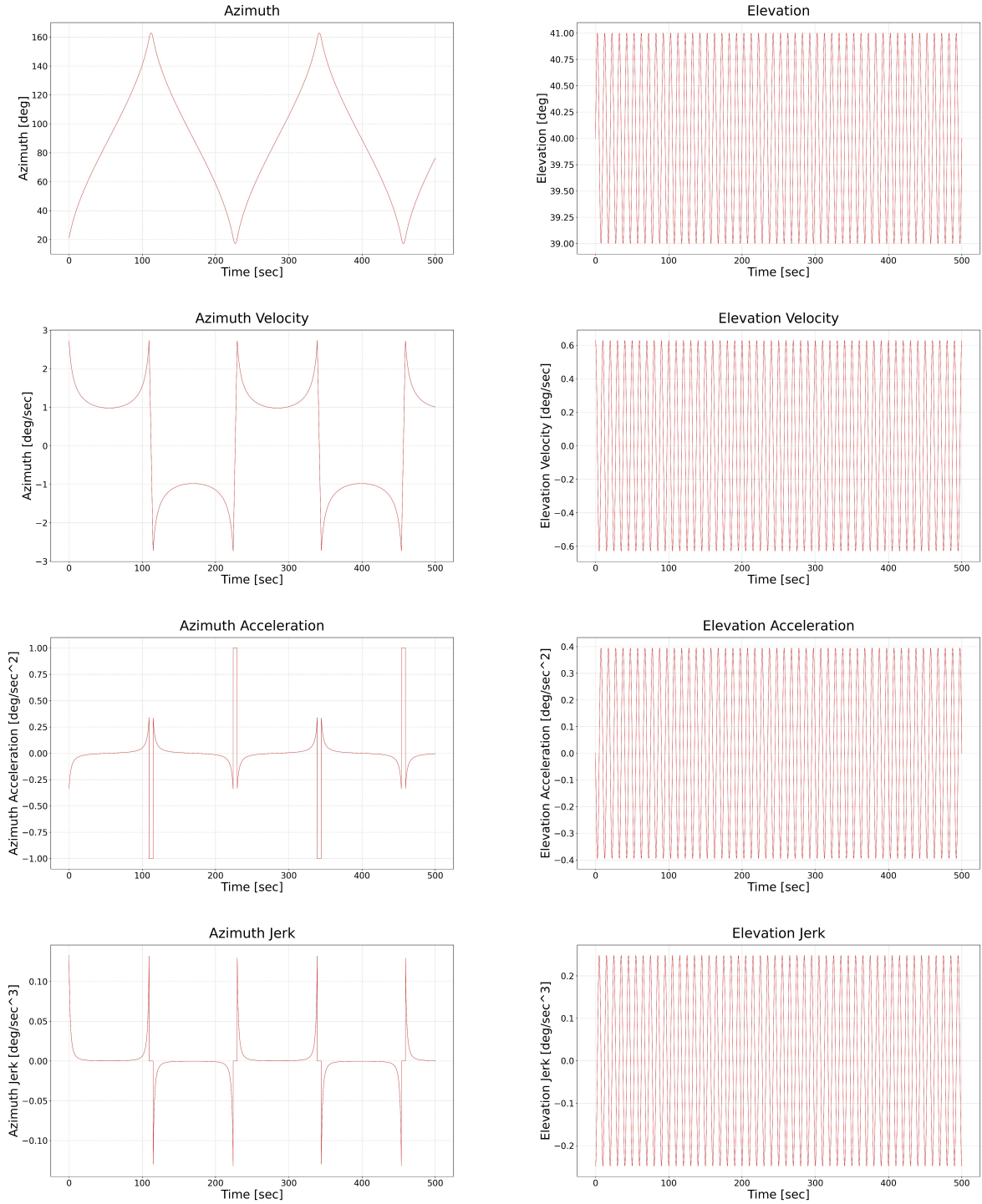

FIG. 13. The position, velocity, acceleration, and jerk in horizontal coordinates of the boresight for a rising sky scan of a Sinusoidal Modulated High Cadence Survey Strategy at 40° elevation, azimuthal range of [21°,159°], 1° sinusoidal amplitude, 10s sinusoidal period, and base scan rate 0.75°/sec. The left column shows azimuth statistics, while the right shows elevation statistics. First row is position, second is velocity, third is acceleration, and fourth is jerk. Note that all azimuth statistics are identical to that of a Modulated High Cadence Survey Strategy at 40° elevation, base scan rate 0.75°/sec, and [21°,159°] azimuthal range.

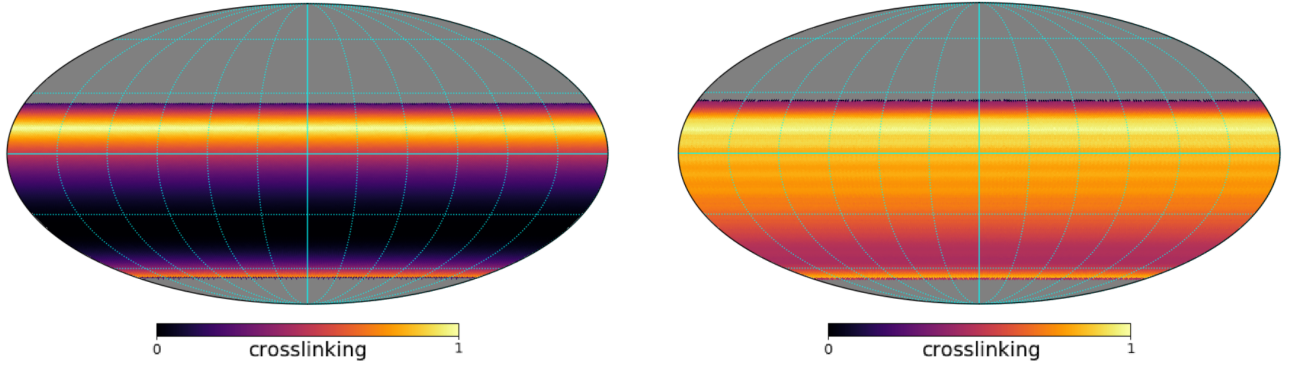

FIG. 14. The crosslinking map of a Modulated High Cadence Survey Strategy (left) and Sinusoidal Modulated High Cadence Survey Strategy (right). Both strategies are at 40° elevation with azimuthal range of [21°,161°], and the sinusoidal amplitude and period for the sinusoidal strategy is 1° and 11s. 0 corresponds to no crosslinking, while 1 corresponds to complete crosslinking. Minimum value for each plot is 0.000 and 0.406, respectively. In Choi et al. (2020) [9] it was found that crosslinking values above 0.3 were preferred for CMB power spectrum analysis. These analyses are part of the motivation for exploring the elevation modulated strategies presented here. Note that the crosslinking parameters used here is defined as unity minus the crosslinking parameter in Choi et al. (2020).

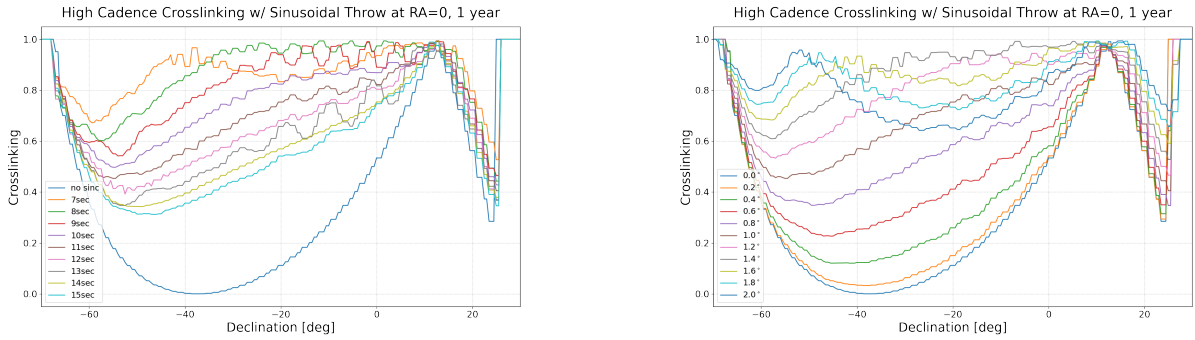

FIG. 15. The values of crosslinking maps of Sinusoidal Modulated High Cadence Survey Strategies with various sinusoidal oscillation periods (left) and amplitudes (right) at 0° right ascension, plotted with respect to declination. 0 corresponds to no crosslinking, while 1 corresponds to complete crosslinking. Note that this is unity minus the crosslinking parameter used in Choi et al. (2020) [9].